\documentclass[aip,preprint,floatfix]{revtex4-1}
\usepackage{times}
\usepackage{graphicx}
\usepackage{psfrag}
\usepackage{epsfig}
\usepackage{ae}
\usepackage{amsmath,amssymb}
\usepackage{booktabs}
\usepackage[version=3]{mhchem} 
\usepackage[usenames]{color}
\usepackage{float}
\usepackage{subfigure}
\usepackage{upgreek}
\draft 

\begin{document}


\title{Structural behavior of a two length scale core-softened fluid in two dimensions} 
 
\author{Daniel Souza Cardoso}
 \affiliation{Instituto Federal Sul Rio-grandense, Rio Grande do Sul, Brasil.}
 \affiliation{Programa de P\'os-Gradua\c c\~ao em  F\'isica, Instituto de F\'isica e Matem\'atica, Universidade Federal de Pelotas. Caixa Postal 354, 96001-970, Pelotas, Brazil.}
 \email{danielcardoso@cavg.ifsul.edu.br}
 \author{Vinicius Fonseca Hernandes}%
\affiliation{Programa de P\'os-Graduac\~ao em  F\'isica, Instituto de F\'isica e Matem\'atica, Universidade Federal de Pelotas. Caixa Postal 354, 96001-970, Pelotas, Brazil.}
 \author{Thiago Pulcinelli Orlandi Nogueira}%
\affiliation{Programa de  P\'os-Graduac\~ao em  F\'isica, Instituto de F\'isica e Matem\'atica, Universidade Federal de Pelotas. Caixa Postal 354, 96001-970, Pelotas, Brazil.}
\author{Jos\'e Rafael Bordin}%
\affiliation{Departamento de  F\'isica, Instituto de F\'isica e Matem\'atica, Universidade Federal de Pelotas. Caixa Postal 354, 96001-970, Pelotas, Brazil.}
 \email{jrbordin@ufpel.edu.br }

\date{\today}

\begin{abstract}
2D competitive systems shows  a large variety of solid and aggregate patterns, re-entrant fluid phase, a continuous melting as predicted by the KTHNY theory and, in some cases, waterlike anomalies. In this paper, we employ Langevin Dynamics simulations of a ramp-like core-softened fluid that have two characteristic length scales and all the features mentioned above. Analyzing the fluid phase of this system, Bordin and Barbosa [Phys. Rev. E 97, 022604 (2018)] reported the existence of two waterlike anomalous region. Now, we revisit this problem focusing in the low temperature regime and a larger range of densities looking for a relation between the origin of the anomalous behavior, the solid phases and the re-entrant melting. Now, not two, but three anomalous regions were observed. The extra anomalous regions are related to the re-entrant melting of a amorphous solid to a cluster fluid. They are ruled not only by the competition between the characteristic length scales in the potential, but also by extra competitions induced by the variety of particles conformations. These competitions extends from the solid to the fluid phase, reflecting in the structural waterlike anomaly. Our results shade some light in the complex behavior of two length scales competitive potential, and helps to elucidate the relation between the large number of solid phases and the existence of more than one waterlike anomalies region in these systems.\end{abstract}

\maketitle 

\section{Introduction}
\label{intro}

In coarse-grained simulations a variety of effective potentials is applied as models for fluids and colloids. Among them, core-softened potentials are a special class employed to understand the behavior of systems with competition. Colloidal suspensions are a good example of such systems, where distinct conformations compete to rule the suspension behavior. Colloids are usually made of molecular subunits which form a central packed agglomeration and a less dense and more entropic peripheral area -- what can be represented by core-softened potentials~\cite{Shukla2008}. The competition in the effective interaction potential can come from the existence of two characteristic length scales in the potential~\cite{Ja98, Ja99a, Ja99b,Oliveira07,Barbosa13, Fomin11, Ya05, Fo08, Lasca10, Buldyrev09} or from softened repulsive potentials~\cite{Saija09, Malescio11, Prestipino10,Prestipino12, Cos13}. As examples, experimental works have shown that the effective interaction in solutions of pure or grafted PEG colloids are well described by core-softened potentials~\cite{colloid1, colloid2, Haddadi20}, and computational studies indicates the same type of effective interactions for polymer-grafted nanoparticles~\cite{Marques20, Lafitte14} or star polymers~\cite{Bos19}.  

This competition between two structures is also observed in the most anomalous system: water~\cite{Angell14, gallo16}. In water, the competition originates from the H bonds breaking and forming between the water molecules. This struggle between two conformations trying to rule the system structure leads to (nearly) 20 known ice phases and more than 70 known anomalies~\cite{URL}. Anomalies are physical-chemical properties whose behavior differ from the observed in most substances. For instance, it is expected that liquids contract upon cooling at constant pressure and diffuse slower upon  compression. However, anomalous fluids expand as the temperature decreases and move faster as the pressure grows. Also, is expected a higher order in the system as the density grows - anomalous fluids does the opposite, they get disordered with the density increase. In this sense, core-softened potentials have been widely used in the literature  to study the  behavior of anomalous fluids~\cite{Ja98, Ja99a, Ja99b, Fomin11, Ya05, Fo08, Lasca10, Prestipino10,Prestipino12, Ma05,Sc00a,Xu05, Ol06a}. Therefore, understanding the behavior of this simple effective models can help to understand the mechanisms behind the behavior of complex fluids as water.

Going from 3D to 2D lead to interesting properties. Since the seminal work by Kosterlitz and Thouless about transition in superfluid films~\cite{Kosterlitz73}, the possibility of unconventional melting scenarios has been extensively studied~\cite{Ryzhov17}. Is this sense, the Kosterlitz-Thouless-Halperin-Nelson-Young(KTHNY) theory showed the existence of a continuous two-stage melting in 2D systems, with a hexatic phase between the solid and isotropic liquids phases~\cite{Gasser10}. This melting scenario was observed for soft-core fluids~\cite{Kapfer15, Rosales06, Zanghellini05}. In fact, competitive core-softened potentials are like a all-in-one 2D interesting phenomena: many studies have observed the anomalous melting scenario with the hexatic phase~\cite{Prestipino12, Dudalov14, Tsiok18, Padilha20}, re-entrant fluid phase~\cite{Lasca10, Prestipino12, Bordin18b, Zhu19, Padilha20} and a large variety of ordered solid phases patterns~\cite{Mendonza09, Patta15,Bos19, Patta17, Scho16,Zhao12,Ciach17, Bordin18a}. And waterlike anomalies as well~\cite{Lasca10, Prestipino12, Camp03, Bordin18b}. Recent studies have been dedicated to understand the relation between these anomalous behaviors~\cite{Lasca10, Prestipino12, Bordin18b}.  Dudalov and co-workers observed that the melting and anomalous behavior of a core-softened system changes going from 3D to 2D~\cite{Duda14}. Krott, Bordin and Barbosa have observed that a core-softened fluid can have two dynamical anomaly regions~\cite{Krott13b}, two density anomalous region~\cite{BoK15a} and an anomalous melting scenario~\cite{Bordin14a} relate to the transition from three to two layers in the quasi-2D limit of strongly confined core-softened fluids. More recently Fomin, Ryzhov and Tsiok~\cite{Fomin20} showed that there is a relation in the freezing temperature and the density anomaly in the quasi-2D limit. In the 2D case, Bordin and Barbosa observed not one, but two regions with waterlike density, dynamics and structural anomalies in a core-softened potential with two length scales~\cite{Bordin18b}. The first region, at low densities, is related to the competition between the potential characteristic distances, indicated by change in the occupancy in the first two peaks of the radial distribution function. On the other hand, the second anomalous region is located near to a re-entrant melting region. Here, the mechanism behind the anomaly is not the competition between the potential scales, but is related to a ordered-disordered transition in the fluid phase. Nevertheless, despite the large number of works in these systems, there are still some questions opened about the connection between all of these behaviors.

In this way, we extend our previous work to see the relations between the solid phases, the melting and re-entrant melting, the extra anomalous regions and the length scales in the core-softened potential. As we will show, there are more than only two characteristic distances. If two characteristic distances lead to one anomalous region, extra length scales results in extra anomalous regions~\cite{Rizatti18}. Also, we show that the solid ordering reflects in a cluster fluid, that preserve the solid geometry and is related to the structural anomaly, that spams from fluid to the solid phase. The paper is organized as follow. In the Section~\ref{model} we show the core-softened model employed, the simulation details and the quantities measured to analyze the system. Next, we present our results and their discussion in the section~\ref{result}, followed by our conclusion and perspectives in the Section~\ref{conclu}.

\section{The Model and Simulation Details}
\label{model}

\subsection{The Model}
\begin{figure}[htp]
\begin{center}
\includegraphics[width=0.4\textwidth]{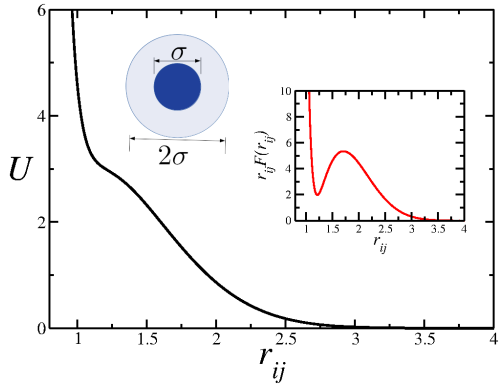}
\end{center}
\caption{Core-softened interaction potential $U$ between two core-corona 
particles. Inset: schematic depiction of the particles, with
the core (first length scale at $r_{ij}\equiv r_1\approx 1.2 \sigma$) and 
the soft corona (second length scale at $r_{ij}\equiv r_2\approx 2.0 \sigma$).}
\label{fig1}
\end{figure}

In this work all the quantities are computed and presented in the standard Lennard Jones (LJ) reduced units~\cite{AllenTild}. Our system consists of $N = 2000$ core-softened disks with a hard core with diameter $r_1$ and a soft corona with diameter $r_2$, and mass $m = 1.0$. It was modeled by a potential interaction composed of a short-range attractive Lennard Jones potential and a long-range repulsive Gaussian barrier centered in $r_0$, with depth $u_0$ and width $c_0$, 
 $$
 U(r_{ij}) = 4\epsilon\left[ \left(\frac{\sigma}{r_{ij}}\right)^{12} -
 \left(\frac{\sigma}{r_{ij}}\right)^6 \right] + 
 $$
 \begin{equation}
 u_0 {\rm{exp}}\left[-\frac{1}{c_0^2}\left(\frac{r_{ij}-r_0}{\sigma}\right)^2\right]\;,
 \label{AlanEq}
 \end{equation}
\noindent where $r_{ij} = |\vec r_i - \vec r_j|$ is the distance  between two disks $i$ and $j$. Depending on the Gaussian parameters, distinct shapes can be obtained with this potential~\cite{Ol06a, Bordin18a}. Here we will study the solid phases in the ramp-like case, which can be obtained with the parameters $u_0 = 5.0$, $c = 1.0$ and $r_0/\sigma = 0.7$. Systems modeled by this potential, showed in figure~\ref{fig1}, exhibits waterlike anomalies for the 3D~\cite{Ol06a, Ol06b}, quasi-2D~\cite{Krott13,BoK15a} and 2D~\cite{Bordin18b} cases. This ramp-like shape lead to  the two competitive length scales in the potential. The short scale is located in  $r_{ij}\equiv r_1\approx 1.2 \sigma$, where the force has a local minimum, and the long range repulsive scale $r_{ij}\equiv r_2 \approx 2 \sigma$, where the fraction of imaginary modes of the instantaneous normal modes spectra has a local minimum~\cite{Charu10}. The cutoff radius for the interaction is $r_c = 3.5$.

\subsection{The Simulation details}

$NVT$ constant molecular dynamics simulations were performed using the Espresso package~\cite{espresso1, espresso2}. The temperature was varied from $T = 0.01$ up to $T = 0.15$. It was kept fixed with the Langevin thermostat~\cite{AllenTild} with $\gamma = 1.0$. The system area,  $A= L^2$, is related to the colloid number density $\rho = N/A$. The density was varied from $\rho = 0.025$ up to $\rho = 0.80$. Standard periodic boundary conditions were employed for the $x$- and $y$-directions.

We performed $3\times10^7$ steps to equilibrate the system. These steps are then followed by $5\times10^7$ steps for the results production stage. The time step was $\delta t = 0.01$, and the equations of motion were integrated using the velocity Verlet algorithm~\cite{frenkelsmit}. To ensure that the system is in equilibrium, the pressure, kinetic and potential energy as function of time were analyzed. The specific heat at constant volume, $c_V$ was obtained by the energy per particle variation along a isochore~\cite{AllenTild}. System snapshots were taken at each $5\times10^4$ steps and analyzed to check structural changes over time. The initial state was created with random position and velocities, and three distinct configurations were employed to obtain the properties. To check for finite size effects, we performed simulations with 6000 particles at random points of the phase diagram. The results showed to be independent from the initial configuration and the number of particles. Is important to address that to describe with precision the melting scenario we should run simulations with a much higher number of particles. In this way, we discuss our results under the light of the KTHNY theory but we must be clear for the reader that for a precise quantitative description of the melting scenario larger simulations would be necessary. However, our findings are in agreement with the prediction of this theory and provide the depiction of this system melting. 

Once the interactions are pairwise important quantities can be calculated explicitly as integrals involving the radial distribution function (RDF) $g(r_{ij})$~\cite{Hansen}. An useful quantity, the two-body contribution to the entropy (or simply two-body entropy $s_2$), can be directly calculated from~\cite{Baranyai89}
\begin{equation}
\label{s2}
s_2 = -\frac{\rho}{2}\int[g(r_{ij})ln(g(r_{ij})) - g(r_{ij}) +1]dr\;.
\end{equation}
It is related to the structural anomaly observed in core-softened systems~\cite{Sh06, Lasca10,BoK15a}: in normal fluids, it decrease under compression, while a increase indicates the anomalous behavior. To check for long range translational ordering using the RDF as basis we evaluate the pair correlation function $h(r_{ij}) = |g(r_{ij}-1|$ and the cumulative two-body entropy~\cite{Klumov20} 
\begin{equation}
\label{cs2}
C_{s2}(r_{ij}) = -\pi \int_0^R [g(r_{ij})ln(g(r_{ij})) - g(r_{ij}) +1]r_{ij} dr_{ij}\;.
\end{equation}
Here $R$ is the upper integration limit. For this work we used $R = 15.0$. At this distance $C_{s2}$ converges for the fluid and amorphous solid phases and diverges for the ordered solid phases. Is important to address that the two-body excess entropy is a structural order metric which connects thermodynamics and structure, not a thermodynamic property of the system. Then it was employed to analyze structural characteristics of the core-softened system, as the waterlike structural anomaly.

While the RDF provides insights on the translational order, the orientational order is checked using the bond orientational order parameter $\Psi_l$,

\begin{equation}
\label{ppsi6}
\Psi_l = \frac{1}{N} \sum_{m=1}^N \psi_l(r_m)
\end{equation}
\noindent where

\begin{equation}
\label{psi6}
\psi_l(r_{m}) = \frac{1}{n_N} \sum_{n=1}^{n_N} \exp[li\theta_{mn}]\;.
\end{equation}
is the local bond orientational order parameter. The sum $n$ is over all the $n_N$ nearest neighbors of $m$ - the neighboring particles were picked by Voronoi tesselation~\cite{Prestipino12}. $\theta_{mn}$ is the angle between some fixed axis and the bond joining the $m-th$ particle to the $n-th$ neighboring particle. For a triangular lattice, $l = 6$ and $|\Psi_6| \rightarrow 1.0 $  if the colloids are in a perfect triangular lattice, and vanishes as it melts. Similarly, $\Psi_4$ is related to the square lattice -- however, this structure was not observed in our simulations. Meanwhile, we observe the stripe phase, a typical structure in simulations of core-softened fluids and experiments for colloidal films~\cite{Fomin19}. In this way, we consider the case $l = 2$ to analyze the twofold stripe order, as proposed by Hurley and Singer~\cite{Hurley92}. As well, we evaluated the orientational correlation
\begin{equation}
    g_l(\vec r) = \langle \psi_l(\vec r)\psi_l^*(\vec 0)\rangle
\end{equation}
\noindent to analyze the long range orientational ordering. The orientational analysis utilized the Freud library~\cite{freud2020}. The cluster size in the fluid phase was analyzed based in the inter particle bonding~\cite{Toledano09,Bordin18a}. Briefly stated, two colloids belong to the same cluster if the distance between them is smaller than the cutoff 1.25 - a value between slightly bigger than the first length scale.

\section{Results and Discussions}
\label{result}

\begin{figure}[htp]
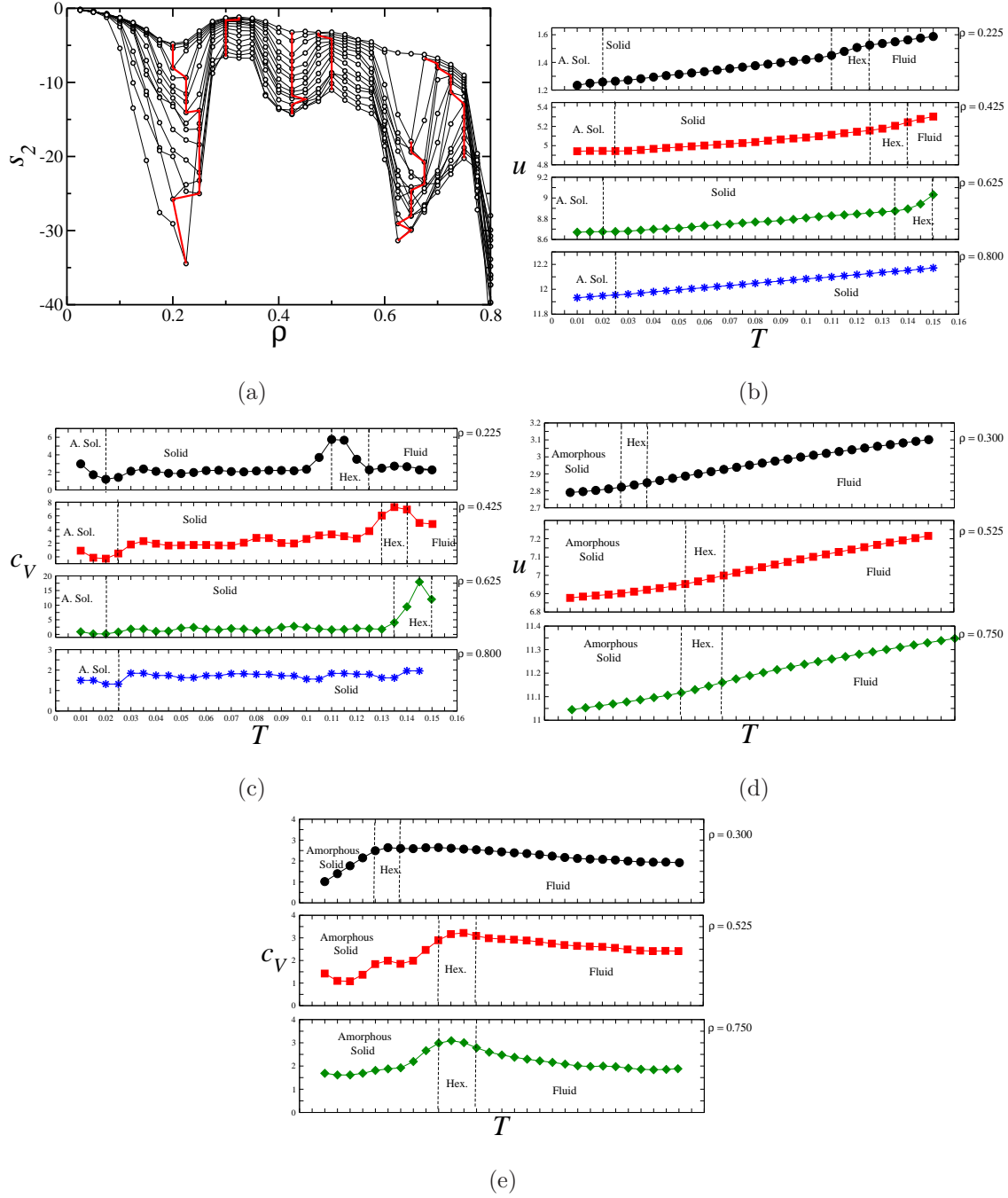

\centering
\subfigure[]{\includegraphics[width=0.45\textwidth]{figuras/fig2a.eps}}
\subfigure[]{\includegraphics[width=0.45\textwidth]{figuras/fig2b.eps}}
\subfigure[]{\includegraphics[width=0.45\textwidth]{figuras/fig2c.eps}}
\subfigure[]{\includegraphics[width=0.45\textwidth]{figuras/fig2d.eps}}
\subfigure[]{\includegraphics[width=0.45\textwidth]{figuras/fig2e.eps}}
\caption{(a) Pair excess entropy $s_2$ as function of the system density for isotherms ranging from $T = 0.01$ to $T = 0.15$. The solid red curves are the minimums and maximums in the curves. (b) Energy and (c) specific heat isocoric curves for densities that crosses the a solid well defined phases: LDT ($\rho = 0.225$), stripe ($\rho = 0.425$), kagome ($\rho = 0.625$) and HDT ($\rho = 0.800$). (d) Energy and (e) specific heat for densities that melts from amorphous solid to fluid clusters with distinct aggregate sizes. }
    \label{fig2}
\end{figure}

We start our discussion with the structural anomaly observed in our system. As we show in the figure~\ref{fig2}(a), the two-body entropy have an unusual behavior. One expects that entropy decays with the increase of the density, once we expect a higher ordering under higher packing. But here we observe a series of minimums and maximums, indicating a series of ordered/disordered transitions at all temperatures. Three different anomalous regions were observed, as we indicate by the red curves in the figure~\ref{fig2}(a). This result in not surprising - similar behaviour were observed in our previous studies in 2D and quasi-2D systems~\cite{Bordin18b, Krott13b, BoK15a}. The question is how each one of this anomalous region - which starts in the solid, pass by the hexatic phase and enter the fluid phase - are related to re-entrant fluid phases and the transition between distinct solid phases.

\begin{figure}[htp]
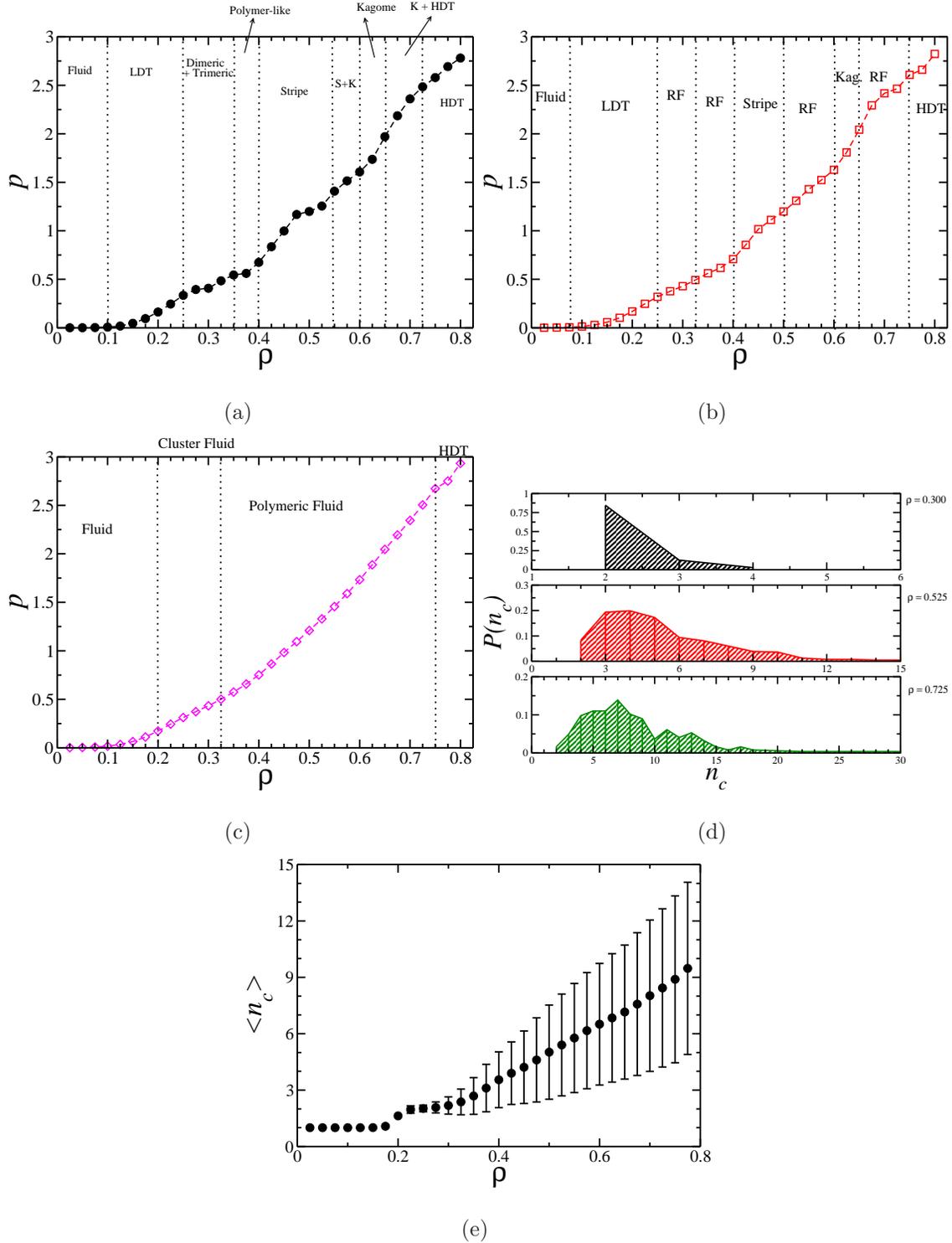

\centering
\subfigure[]{\includegraphics[width=0.45\textwidth]{figuras/fig3a.eps}}
\subfigure[]{\includegraphics[width=0.45\textwidth]{figuras/fig3b.eps}}
\subfigure[]{\includegraphics[width=0.45\textwidth]{figuras/fig3c.eps}}  
\subfigure[]{\includegraphics[width=0.45\textwidth]{figuras/fig3d.eps}} 
\subfigure[]{\includegraphics[width=0.45\textwidth]{figuras/fig3e.eps}} 
\caption{ Pressure vs density isotherms for (a) $T = 0.03$, (b) $T = 0.10$, and (c) $T = 0.15$. Clustering in the fluid phase at $T = 0.15$ indicated by the mean number of particles in each cluster (d) and cluster size distribution (e).}
    \label{fig3}
\end{figure}

2D colloidal systems shows a rich variety of solid phases. Our case is not different. At lower densities the system is frozen in a triangular solid phase - we will call it Low Density Triangular (LDT). Increasing the density it changes to a stripe solid, and then a kagome lattice and finally a High Density Triangular (HDT) phase. In figure~\ref{fig2}(b) and (c) we show the potential energy per particle and the specific heat as function of the temperature for four isochores. The density $\rho = 0.225$ cross the solid LDT phase, density $\rho = 0.425$ the stripe phase, density $\rho = 0.625$ the kagome and the density $\rho = 0.800$ the HDT phase. The curves indicates a smooth solid-fluid transition with a intermediate hexatic phase, in agreement with the KTHNY scenario. As we show further, the hexatic phase have a translational and orientational ordering intermediary between the solid and fluid phases. Also, the energy behavior indicates the existence of other phase transition at low temperatures, that we have identified as an amorphous/disordered solid to ordered solid transition. In fact, for some densities, only this amorphous phase is observed. Examples of these isochores are $\rho = 0.300$, $\rho = 0.525$ and $\rho = 0.725$, whose energy and specific heat are shown in  the figures~\ref{fig2}(d) and (e). The isochores without a well defined solid phase are in the same density region of the extra anomalies regions - the increase of $s_2$ with density - and the re-entrant melting. 

The smoothness of the transition can be observed in the $p\times\rho$ curves as well. Along the isotherm $T = 0.03$, shown in the figure~\ref{fig3}(a), the system is fluid at low densities. We can see a series of transitions related to changes in the curve slope - such changes are the transition from ordered to disordered to ordered solid. Looking for a higher temperature, as $T = 0.10$, the $p\times\rho$ behavior is similar, as we can see in the figure~\ref{fig3}(b). However, here is the re-entrant fluid (RF) phase that separates the ordered solid phases. At higher temperatures, $T = 0.15$, the system is only solid at higher densities, as the $p\times\rho$ curve, shown in figure~\ref{fig3}(c), indicates. However, even here the curve slope changes. This indicates some kind of fluid-fluid transition -- that can be related to a fluid clustering. The mean number of particles in each cluster $<n_c>$, shown in the figure~\ref{fig3}(d), indicates that the clustering start at the density $\rho = 0.200$ with the aggregation in dimers, and above $\rho = 0.300$ $<n_c>$ it grows monotonically with the density. This system is another example of a phase consisting of spontaneously formed dimer in a 2D systems that exhibit a string phase -- something only recently reported in simulations~\cite{Nowack19} and experiments~\cite{Haddadi20}. The huge errors bars in the mean value indicates that there is a large variety of clusters sizes - which is confirmed by the probability to observe a cluster with size $n_c$, $P(n_c)$, shown in figure~\ref{fig3}(e). Along the fluid phase in the isotherm $T = 0.15$, the first minimum in $s_2$
coincides with the free particles to dimer transition, while the first maximum with the dimer to polymer transition. This reflects to the solid phase at $T=0.03$, where the LDT,  whose separation between the particles in the triangular lattice is at the second scale, changes first to a dimeric disordered solid. The dimers grow to polymers and then align in the stripes solid phase. The stripes melts to the re-entrant region, as for $T=0.10$, where we observe the dimeric/polymeric fluid between these ordered solid phases. Similarly, the second anomalous region in the fluid phase is when some of the polymers starts to buckle to ring like structures with smaller triangular subunits - this corresponds, in the solid phase, to a transition from the stripes to the kagome lattice, with a intermediate ring-like amorphous phase. Finally, the third anomalous region - and third re-entrant phase - is related to the triangular fluid rearrangements, a reflection from the kagome to HDT transition with a intermediate amorphous phase with honeycomb-like clusters. Now let's analyze the structural properties that corroborates these findings and elucidate the relation between the waterlike structural anomaly and the solid phases.

\begin{figure}[htp]
\centering
\subfigure[]{\includegraphics[width=0.45\textwidth]{figuras/fig4a.eps}}
\subfigure[]{\includegraphics[width=0.45\textwidth]{figuras/fig4b.eps}}
\subfigure[]{\includegraphics[width=0.45\textwidth]{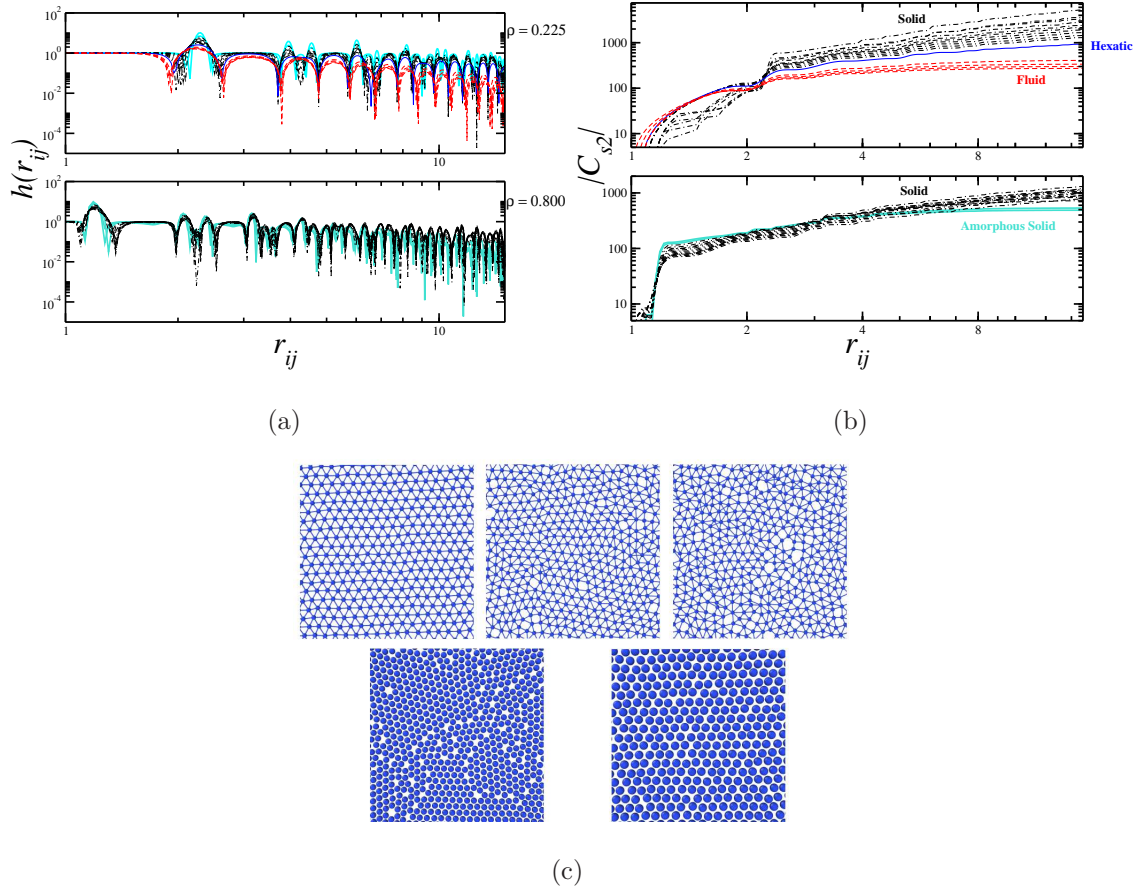}}
\caption{(a) Pair correlation functions $h(r_{ij}) = |g(r_{ij}-1|$ at low density ($\rho = 0.225$) and high density ($\rho = 0.800$) that crosses the triangular phases from $T=0.01$ to $T=0.15$ with a increment $\delta T = 0.01$. Dot-dashed curves are for solid phase, solid blue lines for the hexatic phase and dashed red lines for the fluid phase. (b) Cumulative two-body entropy for distinct temperatures along the isochore. (c) System snapshot from the LDT solid to the fluid phase for $\rho = 0.225$ (upper panel) and for the amorphous and HDT solid phase at $\rho = 0.800$ (lower panel).}
    \label{fig4}
\end{figure}

Once the two-body entropy is related to the structural properties of the system, is important to understand the translational behavior at distinct densities. First, let's analyse the system in isochores at the LDT and HDT phases. The pair correlation functions $h(r_{ij}) = |g(r_{ij}-1|$ are shown in figure~\ref{fig4}(a) for temperatures 0.01, 0.02, 0.03, ..., 0.15. As expected, both solid phases are similar - the difference is that in the LDT phase the particles are at the second scale, $r_2$, and in the HDT phase they are packed at the first length scale $r_1$. For $\rho = 0.225$, a isochore in the LDT phase, the system is in a disordered amorphous solid phase at low temperatures -- turquoise solid lines in the figure, get organized to a triangular solid -- dot-dashed black lines -- and melts to the fluid phase -- dashed red line -- in the simulated range of temperatures, with a hexatic phase -- solid blue line -- between the solid and fluid phases. In the amorphous/LDT and in the amorphous/HDT transition is possible to see a clear change in the pair correlation functions. Also, in the LDT/fluid melting there is a intermediate hexatic curve, whose behavior is between the observed LDT and fluid. Nevertheless, in all phases we can see similarities - indicating that the fluid phase have the same symmetry from the solid phase~\cite{Nowack19}.

To explore deeper the long range translational ordering we analyze the cumulative two-body entropy $C_{s2}$, shown in the figure~\ref{fig4}(b). As we can see in the dot-dashed black curves of the higher panel, when the system is in the LDT solid phase $C_{s2}$ increases with the distance, indicating a long range ordering. In the hexatic phase the slope of the curve changes, as the blue solid curve shows, with an intermediate behavior between the solid and fluid phases. This indicates the existence of ordering, but with small range that the one observed for the solid triangular phase. Finally, the dashed red curves shows that for the fluid phase there is ordering with a range of $\approx 2r_2$. Instantaneous snapshots from these phases are shown in the upper panel of the figure~\ref{fig4}(c), with the bonds indicating the melt from a triangular lattice to a fluid phase with an intermediate hexatic phase. We can see that the triangular symmetry even in the fluid phase. Certainly it is related to the structural anomaly spamming from the solid to the fluid phase. The short range structure is also clear in the amorphous solid phase observed at the lower temperatures in the isochore $\rho = 0.800$, represented by the solid turquoise lines in the lower panel of the figure~\ref{fig4}(b). The shorter ordering is related to defects in the triangular lattice, as the snapshot in the lower panel of figure~\ref{fig4}(c) shows. We can understand this defects by the characteristic of the interaction potential. To reach the perfect triangular crystal lattice at the hard-core distance, $r_1$, is required some kinetic energy for the particle to leave the second length scale, climb up the ramp and reach the first scale. In other words, the entropic contribution for the free energy has to be enough to overcome the entalpic penalty induced by the ramp. Then, upon heating the system can overcome the entalpic barrier, changing from the amorphous solid with short range ordering to the high density triangular phase with long range ordering, as we can see in the lower panels of figure~\ref{fig4}(b) and (c).

\begin{figure}[htp]
\centering
\subfigure[]{\includegraphics[width=0.45\textwidth]{figuras/fig5a.eps}}
\subfigure[]{\includegraphics[width=0.45\textwidth]{figuras/fig5b.eps}}
\subfigure[]{\includegraphics[width=0.45\textwidth]{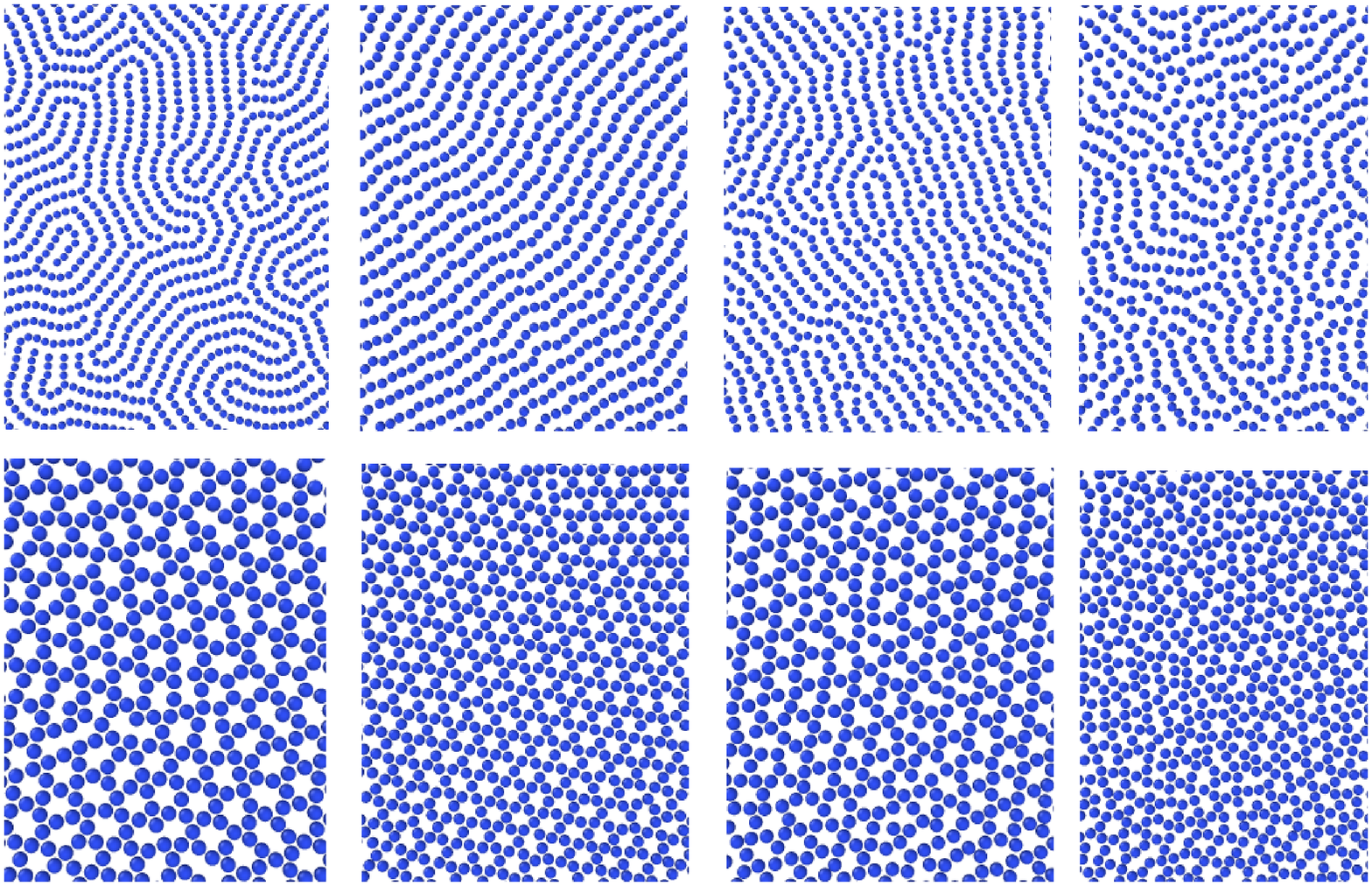}}
\caption{(a) Pair correlation functions $h(r_{ij}) = |g(r_{ij}-1|$ at the stripes region ($\rho = 0.425$) and kagome region ($\rho = 0.625$) from $T=0.01$ to $T=0.15$ with a increment $\delta T = 0.01$. Dot-dashed curves are for solid phase, solid blue lines for the hexatic phase and dashed red lines for the fluid phase. (b) Cumulative two-body entropy for distinct temperatures along the above isochores. (c) System snapshot from the amorphous solid, solid, hexatic and fluid phases for $\rho = 0.425$ (upper panel) and for  $\rho = 0.625$ (lower panel).}
    \label{fig5}
\end{figure}

\begin{figure}[htp]
\centering
\subfigure[]{\includegraphics[width=0.45\textwidth]{figuras/fig6a.eps}}
\subfigure[]{\includegraphics[width=0.45\textwidth]{figuras/fig6b.eps}}
\subfigure[]{\includegraphics[width=0.45\textwidth]{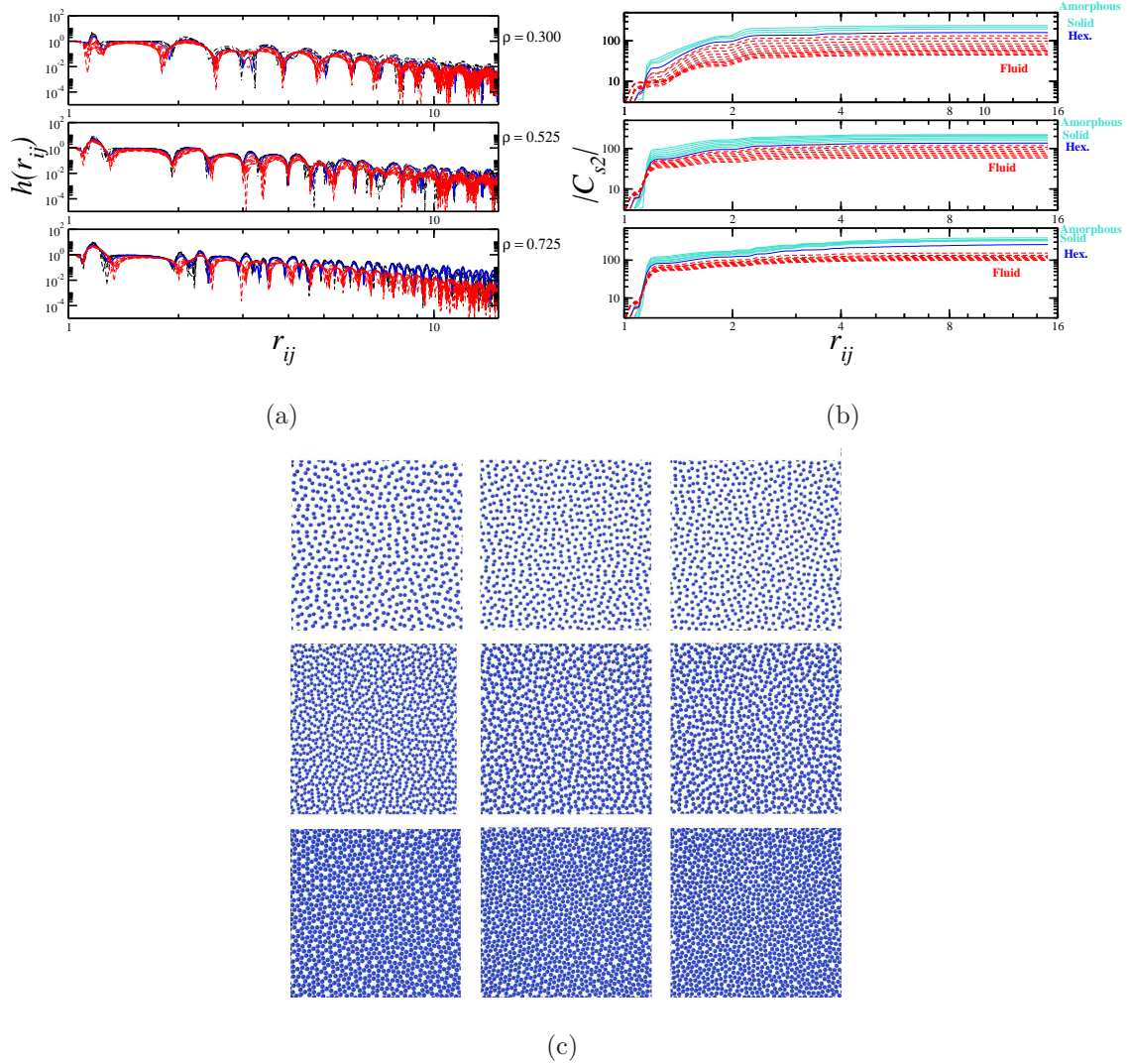}}
\caption{(a) Pair correlation functions $h(r_{ij}) = |g(r_{ij}-1|$ at the amorphous region with re-entrant melting for $\rho = 0.300$, 0.525 and 0.725 from $T=0.01$ to $T=0.15$ with a increment $\delta T = 0.01$. Dot-dashed curves are for solid phase, solid blue lines for the hexatic phase and dashed red lines for the fluid phase. (b) Cumulative two-body entropy for distinct temperatures along the above isochores. (c) System snapshot from the amorphous solid, hexatic and fluid phases for $\rho = 0.300$ (upper panel),for  $\rho = 0.525$ (middle panel) and for  $\rho = 0.725$ (lower panel).}
    \label{fig6}
\end{figure}

For the well defined solid phases at intermediate densities, the stripe and kagome structures, the scenario is similar. As we show in the figure~\ref{fig5}(a), the $h(r_{ij})$ indicates a transition from the solid to the fluid phases with a intermediate haxatic phase. The same conclusions we can get from the $C_{s2}$ behavior in figure~\ref{fig5}(b). They also indicate the existence of an amorphous solid phase. For the stripe region, $\rho = 0.425$, the system is, at low temperatures, an amorphous labyrinth-like solid. Once again, here is the entropy vs entalpy struggle characteristic of competitive systems, in such way that upon heating the labyrinth becomes well defined stripes - as show in the upper panels of figure~\ref{fig5}(b) and (c). Keeping heating up, the stripes starts to break in polymer-like small pieces - in the hexatic phase we can see a coexistence of long stripes and smaller polymer-like clusters, while in the fluid phase we only observe the polymeric clusters. In the kagome region, $\rho = 0.625$, the minimal structure are now trimers with a triangular shape.The amorphous phase, shown in the figure~\ref{fig6}(c), looks like formed by ring-like structures merging together - something between the labyrinth phase and the kagome lattice. In the twist from stripe to ring the triangular cluster is formed. Then, the evolution upon heating is from a disordered solid of triangular trimer, that rearrange to the kagome lattice when they have sufficient entropic contribution to the free energy, and then melts to a triangular trimer cluster fluid with a intermediate hexatic phase -- as show in the lower panels of figure~\ref{fig5}(b) and (c).

As we stated before, there are regions where the system melts from an amorphous solid to a cluster fluid phase that coincides with the re-entrant fluid phases and the fluid clustering. The structural transition from amorphous solid to colloidal fluid phase is shown in figure~\ref{fig6}(a) for densities in the first, second and third re-entrant region.  Here we pick the densities that corresponds to the maximum penetration of the fluid phase - or the isochores with the smaller melting temperature. It is possible to distinguish in the $h(r_{ij})$ at distinct temperatures distinct behaviours. All of them shows a fast convergence for $C_{s2}$, characterizing a short range ordering in the system, as the figure~\ref{fig6}(b) shows. The snapshots in the figure~\ref{fig6}(c) for the three cases indicates that these morphology corresponds to mixtures of distinct clusters geometries. As $\rho$ increases, in the solid phase, the system leaves the LDT phase and rearranges to the stripe phase. However, the aggregation in dimeric clusters leads to the amorphous solid phase, as the upper panel of the figure~\ref{fig6}(c) shows. Similarly, between the stripe and kagome phases there is a mixture of polymer-like and triangular cluster, see the middle panel of the figure~\ref{fig6}(c), and in the region between the kagome and the HDT lattices a mixture of these structures, as shown in the lower panel of the figure~\ref{fig6}(c). 

\begin{figure}[htp]
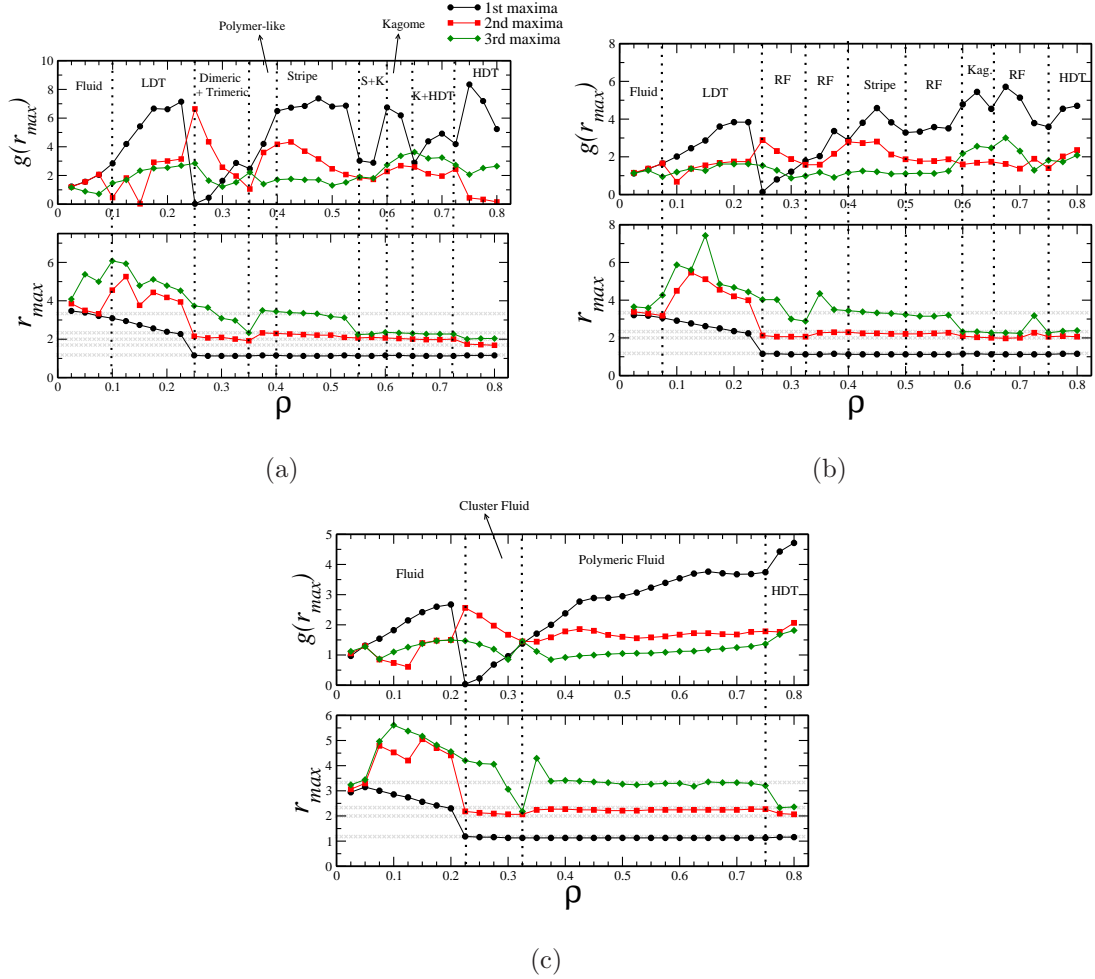

\centering
\subfigure[]{\includegraphics[width=0.45\textwidth]{figuras/fig7a.eps}}
\subfigure[]{\includegraphics[width=0.42\textwidth]{figuras/fig7b.eps}}
\subfigure[]{\includegraphics[width=0.42\textwidth]{figuras/fig7c.eps}}
\caption{A schematic depiction of the length scales is show in (a). Height $g(r_{max})$ and position $r_{max}$ of the first three peaks in the RDFs as function of density along a (b) ordered-disordered solid isotherm ($T = 0.03$), (c) solid/re-entrant melting isotherm ($T = 0.10$) and (d) fluid/cluster fluid/solid isotherm ($T = 0.15$). }
    \label{fig7}
\end{figure}

\begin{figure}[htp]
\centering
\subfigure[]{\includegraphics[width=0.45\textwidth]{figuras/fig8a.eps}}
\subfigure[]{\includegraphics[width=0.45\textwidth]{figuras/fig8b.eps}}
\subfigure[]{\includegraphics[width=0.45\textwidth]{figuras/fig8c.eps}}
\subfigure[]{\includegraphics[width=0.45\textwidth]{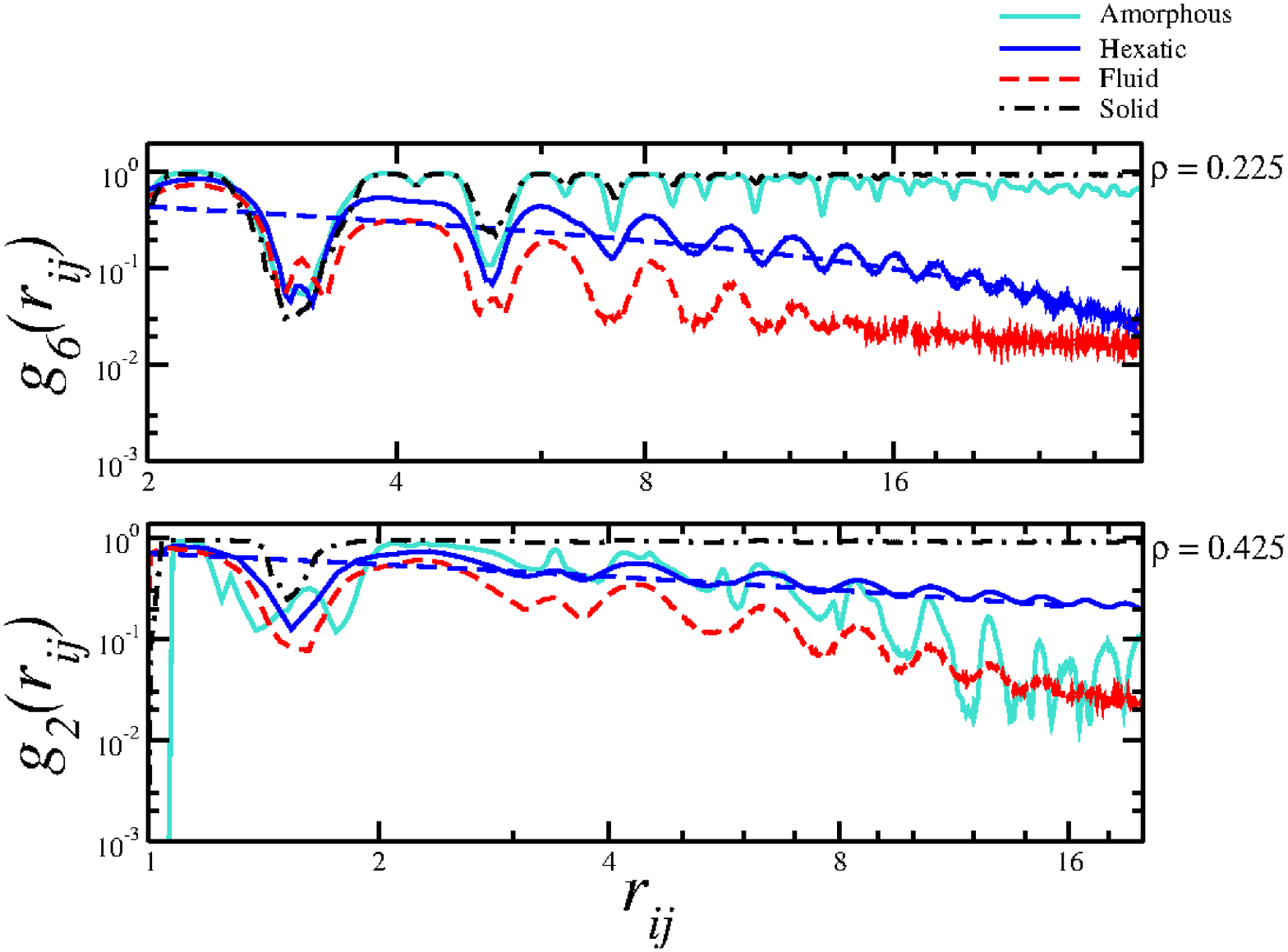}}
\caption{Orientational order parameters (a) $\Psi_6$, (b) $\Psi_4$ and (c) $\Psi_2$ for temperatures from $T=0.01$ to $T=0.15$ with a increment $\delta T = 0.01$. (d) Correlation parameters $g_6$ for a density in the LDT phase ($\rho = 0.225$) and $g_2$ for a density in the stripe phase ($\rho = 0.425$). The blue dashed line is a curve $\approx r_{ij}^{-1/4}$. }
    \label{fig8}
\end{figure}

All this indicates similarities emerging from the amorphous solid phase to the re-entrant fluid phase. To relate it with the competitive structures in the system we can analyze the occupancy in the RDF peaks. We can expect a high occupancy in the potential length scales $r_1$ and $r_2$. Likewise, the sum of these length scales should be relevant. However, the snapshots shows a large variety of clusters and solid patterns. For instance, in the stripe phase the length scale $r_{12}^h$, that corresponds to the hypotenuse of a right triangle with cathetus $r_1$  and $r_2$, is favored. Similarly, in the kagome and HDT phases the length scale $r_{11}^h$, the side of a isosceles triangle with base $r_1$, is relevant. To relate it with the competitions in the system, we evaluate the height $g(r_{max})$ and position $r_{max}$ of the first three maximum in the RDFs. They are shown in the figure~\ref{fig7} for the three isotherms that we are analyzing: (a) $T = 0.03$, (b) $T = 0.10$ and (c) $T = 0.15$. In the $r_{max}$ curves we indicate in grey the length scales. In this same figure we show the behavior for the isotherm $T = 0.03$. At low densities, smaller than $\rho = 0.100$ the system is in the fluid phase. As $rho$ increases, the particles start to packing in the LDT phase. Initially, the separation between the first neighbours (first peak position) is greater than the second scale $r_2$ and approaches to it as the packing increase. At the density $\rho = 0.250$ the system is in an almost perfect triangular lattice. It is indicated by the peak in the orientational order parameter $\Psi_6$, shown in the figure~\ref{fig8}(a) -- $\Psi_6 = 1.0$ for the perfect triangular lattice. As this parameter decay we observe the movement from the first peak from $r_2$ to $r_1$ - the well known competition between the scales. Then the occupancy in the first peak - now at the first scale - decays and the occupancy of the second peak is higher. This is the dimer to polymer-like cluster arrangement - or the first amorphous phase. As the density and the polymer size increases the occupancy at $r_1$ increases as well. Likewise, now the third peak becomes more pronounced than the second one and moves into the direction of $r_2$. The limit when the third and the second maximum merge to a single peak coincides with the rearrangement from the labyrinth like structure to the stripe ordering. The stripe ordering resembles a stretched square lattice. The square symmetry is clear in the $\Psi_4$ behavior, figure~\ref{fig8}(b). But this pattern  can also be related to the orientational order parameter $\Psi_2$ show in the figure~\ref{fig8}(c). At the stripe phase, the new merged second peak leaves $r_2$ and moves to $r^h_{12}$ and the third peak is further now, at $r_1+r_2$. As we see, only the two characteristic length scales from the interaction potential are not enough to understand these structures. The disorder-order transition at higher densities depends on the length scales $r^h_{12}$ and $r_1+r_2$. It is clear when we see the transition from stripe to amorphous: at $\rho = 0.525$ the third peak returns to $r^h_{12}$ and the occupancy at this shell overcomes the occupancy in the second peak as $rho$ increases -- which remains located in $r_2$. This competition between the second and third peaks is responsible for the entrance in the Kagome phase at $\rho = 0.575$. The third peak keeps growing and, at $\rho = 0.650$, where they have the higher occupancy, the system turn out to be disordered. Now, at this high packing the second and third shells position have just a small shift: the second peak moves to $r^h_{11}$ and the third to $r_2$.  So a small change in the RDF second and third peak position lead to completely distinct structures -- in agreement with the conclusion of Nowack and Rice~\cite{Nowack19}, that stated that "the many particle free energy surface is likely very complicated with many minima separated by small barriers, and it is likely that nearby minima correspond to particle packing with different symmetries".


Similar occupancy transitions can be observed in the ordered solid to re-entrant fluid transition, as shown in the upper panel of the figure~\ref{fig7}(a) for $T = 0.10$. As well, a similar jump between the scales at $r_1$, $r_2$, $r_h $ and $r_{12}$ are observed in the borders between the phases. Therefore, the same competition between the scales that are responsible for the order-disorder transition in the solid phase are also responsible for the re-entrant melting line. Even in the fluid phase, $T=0.15$ shown in figure~\ref{fig7}(b), the competition between the proposed scales is observed. However, at this higher temperature, the competition between the preferable positions is related to a transition between a disperse fluid to a cluster fluid, with distinct cluster size and shapes at distinct densities. We can see that the fluid phase preserves not only the solid translational ordering, ruled by the scales, but also the orientational ordering. The orientational correlation parameters $g_6$ for the densities $\rho = 0.225$ and $g_2$ $\rho = 0.425$, showed in the figure~\ref{fig8}(d), indicates that the fluid phase, the red dashed curves, are similar to the hexatic curves - solid blue curves. As expected, the hexatic phase decays with $r_{ij}^{-1/4}$, the ordered solid dot dashed curve converges to a constant and the amorphous solid turquoise curve does not. The fluid orientation is also clear once it does not decay with  $r_{ij}^{-1}$, as we expected for a isotropic fluid. This reinforces the fact that the distinct re-entrant phases and anomalous regions are related to the fluid clustering, and that the clusters preserved the translational and orientational characteristics from the solid phases.

\begin{figure}[htp]
\centering
\includegraphics[width=0.485\textwidth]{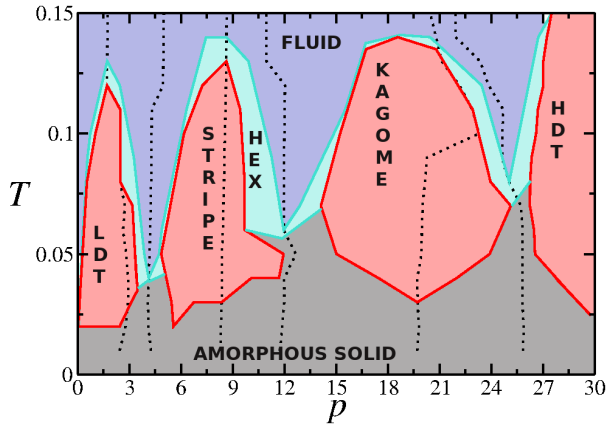}
\caption{$Tp$ phase diagram with the ordered (red) and disordered (gray) solid regions, the hexatic region (turquoise) and the fluid region (blue). Dotted black lines are the minima and maxima in the $s_2$ curve from figure~\ref{fig2}(a).}
    \label{fig9}
\end{figure}

With this, we see that the competition between the same rules the order-disorder transition in the solid phase, the re-entrant melting line and the waterlike anomalies. Interestingly, our findings corroborates the exact solution for 1D systems by Barbosa and co-workers~\cite{Barbosa13, Rizatti18}. They created potentials with multiples length scales, and showed that if the potential has $L$ length scales it will have $L-1$ anomalous regions. Now, we extend the comprehension showing that even potentials with two length scales can lead to more than two anomalous regions once there are multiple competitions between the multiple conformations in these system. Finally, we show in figure~\ref{fig9}(a) the phase diagram with the ordered/disordered solid regions, the hexatic phase, the re-entrant melting region and the maxima and minima in the two-body entropy.

\section{Conclusions and Perspectives}
\label{conclu}

In this paper we analyzed the behavior of a 2D system modeled by a core-softened potential with two characteristic length scales. The phase diagram has a variety of interesting phenomena, as a KTHNY melting scenario, a re-entrant fluid phase, disordered-ordered solid transitions and multiple waterlike anomalous regions. We show that the extra anomalous regions are related to length scales that characterize the multiple conformations patterns. In fact, the same competition between these scales that leads to different solid phases and induces the ordered/disordered transition at low temperatures is also responsible for the re-entrant phase with a cluster fluid and for the waterlike structural anomaly. 

This leads to some questions about the most important material in our planet: do bulk water have more than one anomalous region~\cite{Fomin17} due its liquid polymorphism and the variety of solid phases -- 17 experimentally observed ice structures and many more predicted by computer simulations~\cite{Salz19}? If not in bulk, but in quasi-2D limit: water confined inside narrow nanopores can have extra anomalous regions~\cite{Krott13b, Bordin14a}? These are open questions that arises under the light of this study.

\begin{acknowledgments}
VFH thanks Coordena\c c\~ao de Aperfei\c coamento de Pessoal de N\'ivel Superior (CAPES), Finance Code 001, for the MSc Scholarship. TPON thanks CAPES, Fincance Code 001, for the PhD Scholarship. JRB acknowledge the Brazilian agencies Conselho Nacional de Desenvolvimento Cient\'ifico e Tecnol\'ogico (CNPq) and Funda\c c\~ao de Apoio a Pesquisa do Rio Grande do Sul (FAPERGS) for financial support. JRB is greatly indebted to Marcia C. Barbosa for illuminating discussions.

\end{acknowledgments}

%

\end{document}